\documentclass[aps,reprint,twocolumn,superscriptaddress]{revtex4-1}
\usepackage{mathtools}
\usepackage{bm}
\usepackage{amsmath,amssymb,graphicx}
\usepackage{mathtools}
\usepackage{bm}
\usepackage{color}
\usepackage{braket}
\usepackage{graphicx}
\usepackage{upgreek}
\usepackage[note-name]{notes2bib}

\begin{document}

\title{Coherent virtual absorption of light in microring resonators}

\author{Q. Zhong} 
\affiliation{Department of Physics, Michigan Technological University, Houghton, Michigan, 49931, USA}
\affiliation{Henes Center for Quantum Phenomena, Michigan Technological University, Houghton, Michigan, 49931, USA}

\author{L. Simonson} 
\affiliation{Department of Physics, Michigan Technological University, Houghton, Michigan, 49931, USA}
\affiliation{Henes Center for Quantum Phenomena, Michigan Technological University, Houghton, Michigan, 49931, USA}

\author{T. Kottos}
\affiliation{Wave Transport in Complex Systems Lab,  Department of Physics, Wesleyan University, Middletown, Connecticut 06459, USA}

\author{R. El-Ganainy}
\email[Corresponding author: ]{ganainy@mtu.edu}
\affiliation{Department of Physics, Michigan Technological University, Houghton, Michigan, 49931, USA}
\affiliation{Henes Center for Quantum Phenomena, Michigan Technological University, Houghton, Michigan, 49931, USA}

\begin{abstract}
Light trapping and radiation process from linear reciprocal photonic resonators is one of the fundamental processes in optical science and engineering.  Recently, the concept of coherent virtual absorption (CVA) of light was introduced and investigated for planar and cylindrical optical structures. The key feature of CVA is that by engineering the time-dependence of the excitation waveform, one can temporarily store all the input energy into the optical structure without any leakage.   Here we further explore this novel concept in integrated photonic setups made of microring resonators. By using coupled-mode theory (CMT), we derive an analytical expression for CVA in this platform. This in turn allows us to make the connection with the notion of coherent perfect absorption (CPA) as well as extending our analysis to active resonators (having optical gain). We next provide a physical insight into this process by using a simple model made of cascaded beam splitters. Importantly, we confirm our results using a full-wave analysis of realistic material systems. Finally, we discuss the limitation on the CVA process due to waveform mismatch and nonlinear effects.     
% {\color{red} OCIS code} 
  
\end{abstract}

%\ocis{(140.3490) Lasers, distributed-feedback; (060.2420) Fibers, polarization-maintaining; (060.3735) Fiber Bragg gratings; (060.2370) Fiber optics sensors.}
% REPLACE WITH CORRECT OCIS CODES FOR YOUR ARTICLE
% NOTE: \ocis{} IS ALIASED TO \pacs{} BUT MUST
% FORMAT THE TERMS CORRECTLY FOR EACH JOURNAL

\maketitle

\section{Introduction}
Light-matter interaction can be tailored by engineering different design parameters such as optical refractive index, magnetic permeability, and/or optical nonlinear interactions to just mention a few examples \cite{Saleh2007NDC}. A whole new world of challenges and opportunities for nascent light-matter interaction effects emerges once we exploit the realm of non-Hermitian photonics \cite{ElGanainy2007P, Musslimani2008O, ElGanainy2008B, ElGanainy2008OB}. Representative examples include a new family of magnetless isolators \cite{Ramezani2010UNI, Aleahmad2017IMP}, microlasers \cite{Hodaei2014PTS, Feng2014SML, kottos2014PRL, Peng2016CM, Maio2016OAM, Zhao2018THS, ElGanainy2015SLA, Teimourpour2016NHE, Hokmabadi2019SLA, Midya2019SMLA}, optical sensors \cite{Wiersig2014ESF, Wiersig2016SOEP, Chen2017EPHS, HosseinESH, Zhong2019SES}, optomechanical devices \cite{Jing2014SPL, Jing2015OIT, Jing2017HOEP, Schonleber2016OI, Xu2016TET, Zhang2018PLO}, etc. For recent reviews, see \cite{ElGanainy2018PRL, Feng2017JMP, Ozdemir2019JPL, Miri2019JPA}. Along these lines, non-Hermitian optics led us to the concept of coherent perfect absorption (CPA), associated with judicious engineering of the spatial waveform of an incident wave in a lossy optical target such that a complete absorption can be achieved \cite{Chong2010JPA}. In its original conception, the CPA was viewed as the time-reversal phenomenon of a laser \cite{Longhi2010JMP, Wan2011JMP,Baranov2017PMJ,Wei2018PRA}. It turns out that the phenomenon is applicable also for systems without any time-reversal symmetry \cite{Li2017PRL} or even more surprising in the absence of any lossy elements \cite{Baranov2017PMK}. In the latter case, it was shown that if the time dependence of the incident waveform is also tailored, one can temporarily store all the incident energy in the structure without any leakage. However, due to the absence of real absorption, all the stored energy will be eventually released as the excitation is turned off. This process, which was called coherent virtual absorption or CVA \bibnote{The terminology 'virtual' should not be confused with the notion of virtual photons.} has been theoretically illustrated for planar and cylindrical optical structure \cite{Baranov2017PMK}; and experimentally demonstrated for mechanical waves \cite{Trainiti2019OOP}. 

In this work, we explore the notion of CVA in a different setting of integrated photonic circuits made of microring resonators. This platform is relevant to technological applications in optical communication systems and quantum information. Interestingly, our analysis based on coupled mode formalism extends the notion of CVA to active systems that exhibit gain. In other words, counter-intuitively, one can temporarily store all the incident energy in the microresonators even when the system should work as an amplifier. We have also investigated the feasibility of achieving CVA in microring resonators using full-wave analysis and realistic material systems. Finally, we investigate the limitations of the CVA protocols in the presence of slight waveform flaws and nonlinear effects. Our results indicate that this scheme is fragile and requires fine tuning of the input signal parameters.

\section{System description}

Let us consider a microring resonator coupled to a waveguide as shown in Fig. \ref{Fig_Schematic}. At each resonant frequency, the ring structure supports two degenerate clockwise (CW) and counterclockwise (CCW) propagating modes. These modes can be externally excited via the two end ports of the waveguide. Thus, the system consists of two input and two output ports. However, due to the wave nature of the traveling modes in the microring resonators, the CW/CCW modes are excited separately. An implicit assumption in our analysis is that the surface roughness of the ring is kept very small \cite{Vahala2005NDC} to avoid any appreciable back reflection between these two modes. Indeed recent state-of-the-art experimental results show that this roughness can be reduced to less than 0.5 nm \cite{Ji2017Optica}.  With this assumption in mind, one can consider only one input port and one output port as shown in Fig. \ref{Fig_Schematic}. Under these conditions, standard temporal coupled-mode theory (TCMT) \cite{Fan2003OOP, Vahala2005NDC} gives:

\begin{figure} [t]
	\includegraphics[width=3.4in]{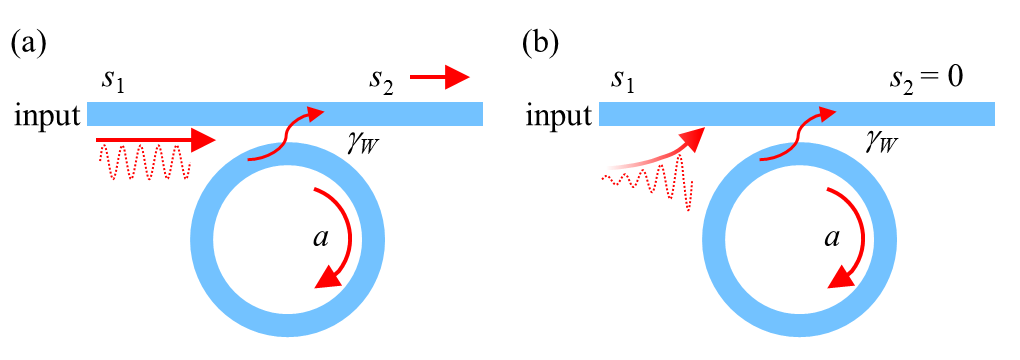}
	\caption{Schematic of a microring resonator coupled to a waveguide. (a) Under constant excitation from the input port, a steady state solution with a finite output is reached. (b) As discussed in detail in this work, by properly engineering the temporal waveform of the input, it is possible to store all the input energy in the ring with nearly zero output.}
	\label{Fig_Schematic} 
\end{figure}

\begin{equation} \label{Eq:CMT}
\begin{aligned}
 \frac{da}{dt} &=(i \omega_0 + g -\gamma_W -\gamma_R-\gamma_A) a + i  r \alpha_0 |a|^2 a + \sqrt{2 \gamma_W} s_1 , 
 \\
 s_2 &=s_1- \sqrt{2 \gamma_W} a,
\end{aligned}
\end{equation} 
where $a$ is the field amplitude of the CW mode, and $s_1$ and $s_2$ quantify the input/output optical signal. In Eq. (\ref{Eq:CMT}), $\omega_0$ is the resonant frequency, $g$ is the optical gain, and $\gamma_{W,R,A}$ are the losses due to the coupling to the waveguide, radiation, and material absorption, respectively. Finally, $\alpha_0$ is the effective nonlinear Kerr coefficient, and the variable $r$ is used as a knob for controlling the strength of the nonlinear interactions (or equivalently the input power levels) as we will see later. 

\section{Results}
First, we consider the case when $r=0$. Under this condition, the lasing threshold of this system is $g_\text{th}=\gamma_W+\gamma_R+\gamma_A$. Below this gain value, when $\gamma_R+\gamma_A<g<g_\text{th}$, the structure functions as an optical amplifier for the input signal. We are particularly interested in that latter regime. Note that the steady state transmission coefficient of the system for an input $s_1=s_0 e^{i \omega t}$ reads $T\equiv\dfrac{s_2}{s_1}=\dfrac{i (\omega-\omega_0)-\tilde{g}-\gamma_W}{i(\omega-\omega_0)-\tilde{g}+\gamma_W}$, where $\tilde{g}=g-\gamma_R-\gamma_A$. The expression for $T$ has a zero at $\omega_\text{zero}=\omega_0 - i (\tilde{g} + \gamma_W)$ and a pole at $\omega_\text{pole}=\omega_0 - i (\tilde{g} - \gamma_W)$, with $\omega_\text{zero} \ne \omega_\text{pole}^*$. Thus, when $\tilde{g}+\gamma_W \neq 0$, the zero transmission regime can be accessed only by using a complex frequency for the external optical input. In practice, this means that the waveform is not just oscillatory but also vary in amplitude. In the particular example studied here, the input signal must have a central frequency $\omega_0$ and amplitude  growth rate equal to $\tilde{g}+\gamma_W$, i.e. $s_1=s_0 e^{i \omega_0 t} e^{(\tilde{g}+\gamma_W) t}$. To clearly illustrate this effect and also elaborate on the transient response in the current microring system, we consider the aforementioned excitation and assume that it is switched on at time $t=0$, i.e. $s_1=s_0 e^{i \omega_0 t} e^{(\tilde{g}+\gamma_W) t} u(t)$, where $u(t)$ is the standard unit step function. By solving Eq. (\ref{Eq:CMT}) using the Laplace transformation, we obtain: 

\begin{equation} \label{Eq.sol}
\begin{aligned}
& a(t)=\sqrt{\frac{2}{\gamma}} s_0 e^{i \omega_0 t} e^{\tilde{g}t} \sinh (\gamma_W t) u(t), \\
& s_2 =s_0 e^{i \omega_0 t} e^{(\tilde{g}-\gamma_W) t} u(t).
\end{aligned}
\end{equation} 

Equation (\ref{Eq.sol}) shows that as $t \to \infty$, the optical intensity inside the resonator builds up exponentially ($|a(t)|^2 \propto e^{2 (\tilde{g}+\gamma_W) t}$), while simultaneously the output signal power decays exponentially at a rate equal to $2(\gamma_W-\tilde{g})$. In other words, by engineering the optical amplitude, it is possible to store most of the input energy inside the resonator without light leaking  to the output channel. Thus, when the radiation and material losses are negligible, input power trapping in the resonator becomes nearly perfect even though the system operates in the amplification regime. As a side note, we observe that the decay rate of the output signal in $s_2$ becomes slower as we approach the lasing condition. Specifically, at lasing, the outgoing signal is a propagating wave as expected.  

\begin{figure} [b]
	\includegraphics[width=3.4in]{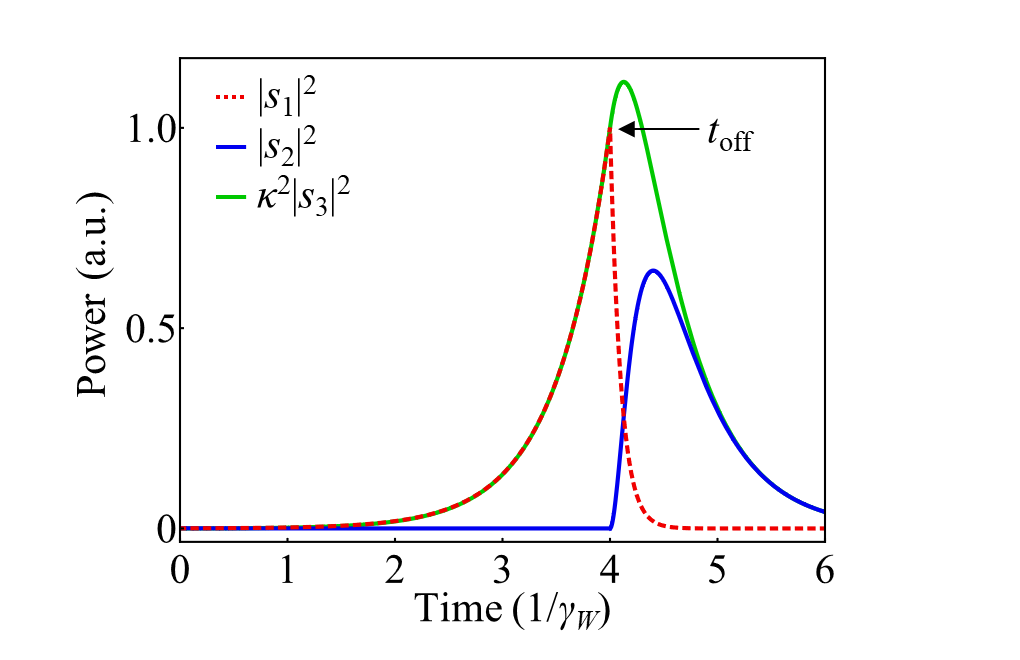}
	\caption{A plot of input/output powers $|s_{1,2}|^2$ (red and blue curves) as well as the scaled power inside the ring $\kappa^2 |s_3|^2=2 \gamma_W |a|^2$ (red curve). The input signal varies exponentially as $e^{\gamma_W t}$ starting from a certain initial value. At $t_\text{off}=4 \gamma_W^{-1}$ the input is switched off with an exponential decay rate of $5\gamma_W$. In plotting these curves,  we normalized all the power values with respect to the maximum input power, i.e. $|s_1|^2$ at $t_\text{off}$ (note that the actual values are not relevant here since we have so far neglected all nonlinear effects). As discussed in the text in detail, the optical power flow is consistent with the analytical expression in Eq. (\ref{Eq.sol}). Note that in the above plot, the value of $\kappa^2 |s_3|^2$ can be larger than unity due to the energy accumulation in the microring resonator.}
	\label{Fig_CMT} 
\end{figure} 

To illustrate these results, we present the numerical integration of Eq. (\ref{Eq:CMT}) in Fig. \ref{Fig_CMT} and we focus first on the passive scenario, i.e. $g=\gamma_R=\gamma_A=0$ under linear conditions with $r=0$. In these simulations, the input signal $s_1$ varies as $e^{\gamma_W t}$ from $t=0$ to $t_\text{off}=4\gamma_W^{-1}$ (the actual value of the input signal at $t=0$ is irrelevant here since the nonlinear effects are neglected). At $t_\text{off}$, it is switched off quickly according to the fast exponential decay $e^{-5 \gamma_W t}$. The power flow in the ring is given by the expression  $|s_3|^2 \equiv|a|^2/\tau$ \cite{Little1997MRC, Heebner2008OM, Vahala2005NDC},
where $\tau=2 \pi R /v_g$ is the round trip time with $R$ and $v_g=c/n_g$ being the ring radius and group velocity of the relevant optical mode at the central frequency $\omega_0$. Here, $c$ and $n_g$ are the speed of light in vacuum and the optical group index, respectively.  To provide an intuitive picture of the simulation results, in Fig. \ref{Fig_CMT}, we plot the scaled power $\kappa^2 |s_3|^2$ (red curve), where  $\kappa=\sqrt{2\gamma_W \tau}$ is the coupling coefficient between the microring waveguide and the external waveguide. Obviously, the quantity $\kappa^2 |s_3|^2=2\gamma_W |a|^2$ represents the power that would flow from the microring to the output port if there was no interference with the input signal. After a transit, and for $t<t_\text{off}$, both $\kappa^2 |s_3|^2$ and the input power $|s_1|^2$ (green curve) become almost identical, with their interference at the output however canceling out and producing nearly zero power as shown for $|s_2|^2$ (blue curve). These results are consistent with the analytical expression of Eq. (\ref{Eq.sol}). On the other hand, for $t>t_\text{off}$, the stored energy is released to the output channel.  

\textbf{Connection to CPA}--- Consider Eq. (\ref{Eq:CMT}) again but with $g=0$ and $\gamma_W = \gamma_R+\gamma_A$. This is the critical coupling condition \cite{Mirza2013JOSAB,Heebner2002JMO,Mirza2019OE,Choi2001OL} which results in steady state zero output (i.e. $s_2|_{t \to \infty}=0$) under a continuous wave excitation $s_1^\text{CPA}=s_0 e^{i \omega_0 t}$, i.e.:
\begin{equation} \label{Eq:CMT_CPA}
\begin{aligned}
& \frac{da^\text{CPA}}{dt} =(i \omega_0 -2\gamma_W) a^\text{CPA} + \sqrt{2\gamma_W} s_1^\text{CPA}, \\
& s_2^\text{CPA}=s_1^\text{CPA}-\sqrt{2\gamma_W} a^\text{CPA}.
\end{aligned}
\end{equation} 
The solution of Eq. (\ref{Eq:CMT_CPA}) is $s_2^\text{CPA}=s_0 e^{i \omega_0 t} e^{-2\gamma_W t}$. 
If we now use the transformation $X=X^\text{CPT} e^{(\tilde{g}+\gamma_W) t}$, where $X \in \{a,s_1 , s_2\}$, we recover Eq. (\ref{Eq:CMT}) with a steady state zero output. This illustrates that the notion of CVA generalizes the concepts of critical coupling and CPA. In words, in the case of CPA based on critical coupling, the optical power is actually absorbed in the system due to the material loss, while in the scenario of CVA, the power is stored inside the resonator due to destructive interference at the output port as we explain in more detail below.

\begin{figure} [t]
	\includegraphics[width=3.4in]{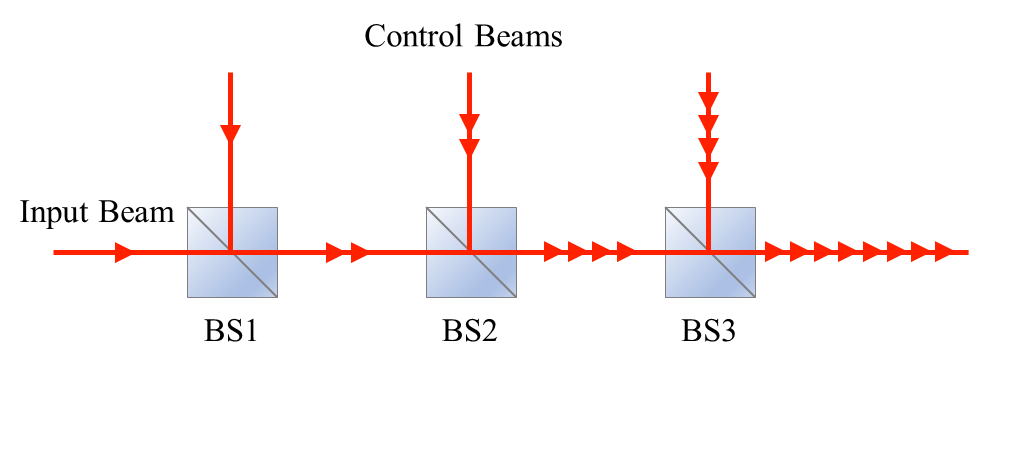}
	\caption{The process of CVA  in a microring resonator can be intuitively understood by considering a cascade of beam splitters (denoted as BS in the figure). At each stage, the amplitude and phase of a control beam (vertical red arrows) can be adjusted in order to interfere constructively with the input beams (horizontal red arrows) in the horizontal direction to arrive at the next stage. To maintain this process at each stage, the amplitude of the control beam at each beam splitter must be larger than that of the previous stage. The junction between the waveguide and the microring is indeed a beam splitter. Input light completing one roundtrip inside the ring meets this junction again which can be considered as another beam splitter stage.}
	\label{Fig_Beamsplitter} 
\end{figure}

\textbf{CVA as a series of beam splitters}--- In order to gain an insight into the process of CVA, let us consider a related problem. The structure shown in Fig. \ref{Fig_Beamsplitter} consists of a series of beam splitters aligned along a horizontal line and we assume that the bulk material between them is transparent (extension to gain medium is straightforward). We also introduce vertical control beams at each beam splitter, whose phases are chosen to interfere constructively with the horizontal input beams along the horizontal output. Thus, the total power of the combined beams will exit from the horizontal direction and travel to the next beam splitter. Clearly, in order to repeat this behavior along all the beam splitters, the amplitudes of the control beams at each stage must be equal to those of the input beam. Thus, at stage $N$, the amplitude of the control beam must be $2^{N-1}$ times that of the original input beam at the first stage, i.e. it must grow exponentially. A direct connection of this cascade process with the CVA occurring at the microring set-up is achieved by realizing that the junction between the waveguide and the microring is indeed a beam splitter. Light entering the resonators travels one roundtrip before it encounters the beam splitter junction again. If after each cycle, both beams have the correct phase and amplitude, they can coherently add in such a way that directs all the energy inside the resonator. This is the essence of the CVA process, which lasts as long as the excitation signal is on. 

\begin{figure} [t]
	\includegraphics[width=3.4in]{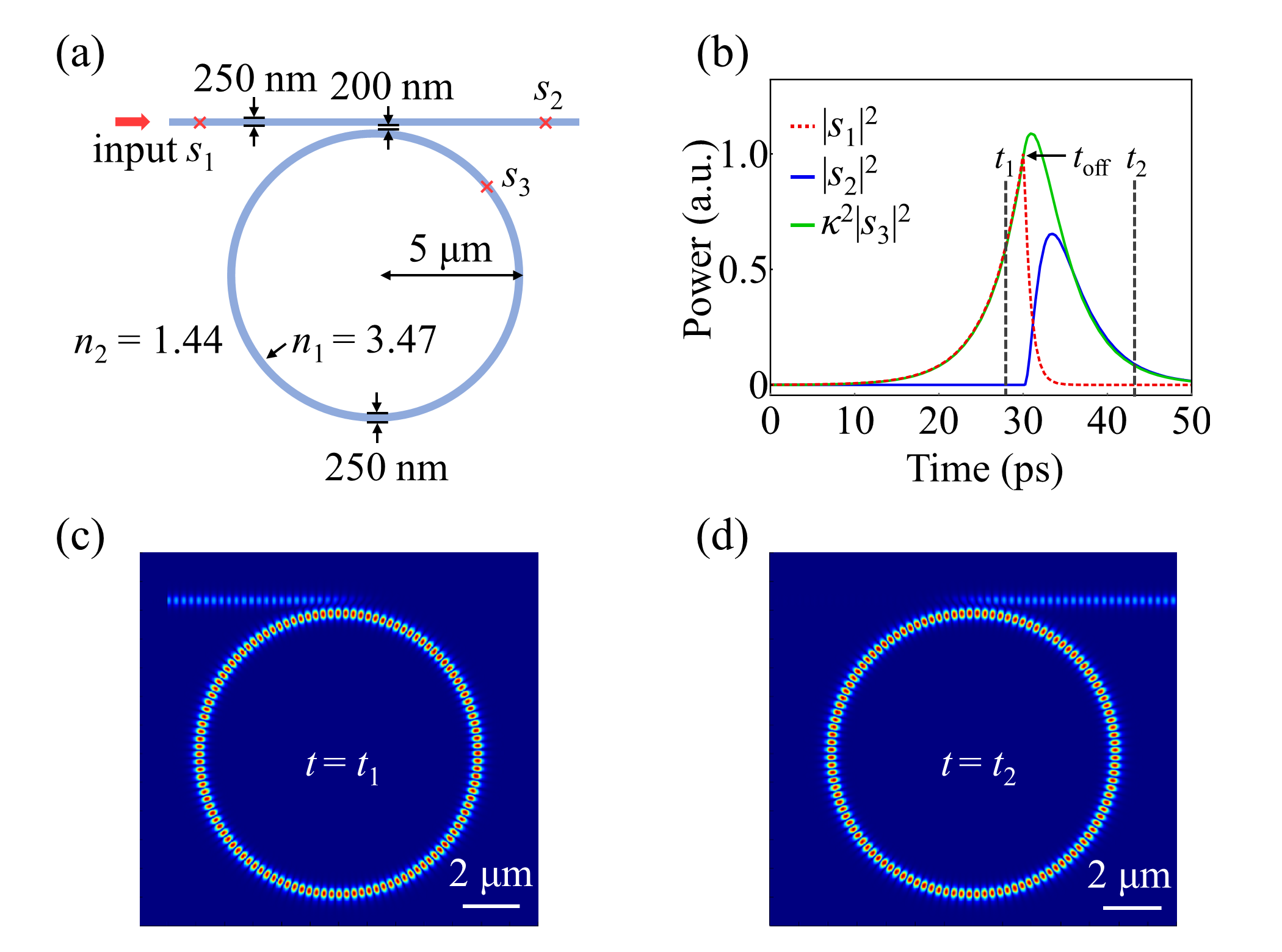}
	\caption{ (a) Schematic diagram of the microring resonator under consideration. The physical dimensions and optical refractive indices are indicated. The results obtained from the simulation of the system. (b) Optical power at the input/output ports as well as the power in the ring (scaled by a factor of $\kappa^2$) when the input power varies exponentially at a rate $\gamma_W$ from time $t=0$ until it is switched off at $t=t_\text{off}$.  The results agree well with those obtained from CMT in Fig. \ref{Fig_CMT}. (c) and (d) are snapshots of electric field distribution in the system at two instants of time, $t=t_{1,2}$ as indicated in (b).}
	\label{Fig_SimulationResults} 
\end{figure}

\textbf{Experimental feasibility}--- Having demonstrated the process of CVA in microring resonators using CMT, we now explore its optical implementation using a realistic material system. To do so, we consider the structure shown in Fig. \ref{Fig_SimulationResults}(a), and we use two-dimensional (2D) full-wave finite difference time domain (FDTD) analysis, which can reveal much of the physics without the high memory and time demand of 3D simulations. The chosen material parameters and dimensions (indicated in the figure) are relevant to silicon photonics platforms \cite{Chrostowski2015NDC}. By employing the aforementioned FDTD scheme, we have estimated that the loss rate from the microring resonator rate to the waveguide at a free space wavelength of  $\lambda_0=1.55 \  \upmu \text{m}$ is $\gamma_W = 0.125 \ \text{ps}^{-1}$. Compared to this value, one can safely neglect the material and radiation loss of the silicon ring resonator at that wavelength. Additionally, in what follows, we also assume that there is no applied gain and neglect the nonlinear effects. Figure \ref{Fig_SimulationResults}(b) plots the FDTD results obtained for the optical Poynting vector (optical power flow) at the input/output ports $|s_{1,2}|^2$ as well as scaled power along the ring waveguide $\kappa^2|s_3|^2$. Here the value of the $\kappa=0.3$ was obtained using FDTD. In obtaining these results, the input signal field was initially exponentially ramped up with a growth rate of $\gamma_W$. At time $t=t_\text{off}=30 \ \text{ps}$, the input signal was switched off exponentially with a fast decay rate of $5\gamma_W$. As can be clearly seen, during the initial phase the optical power is stored in the ring with near zero output. As expected, this power is then released to the output port in the second phase. Furthermore, we plot a snapshot of the electric field distribution along the structure at both the initial and final phases (times $t=t_{1,2}$), which illustrate the energy storage then release. 

\begin{figure} [t]
	\includegraphics[width=3.4in]{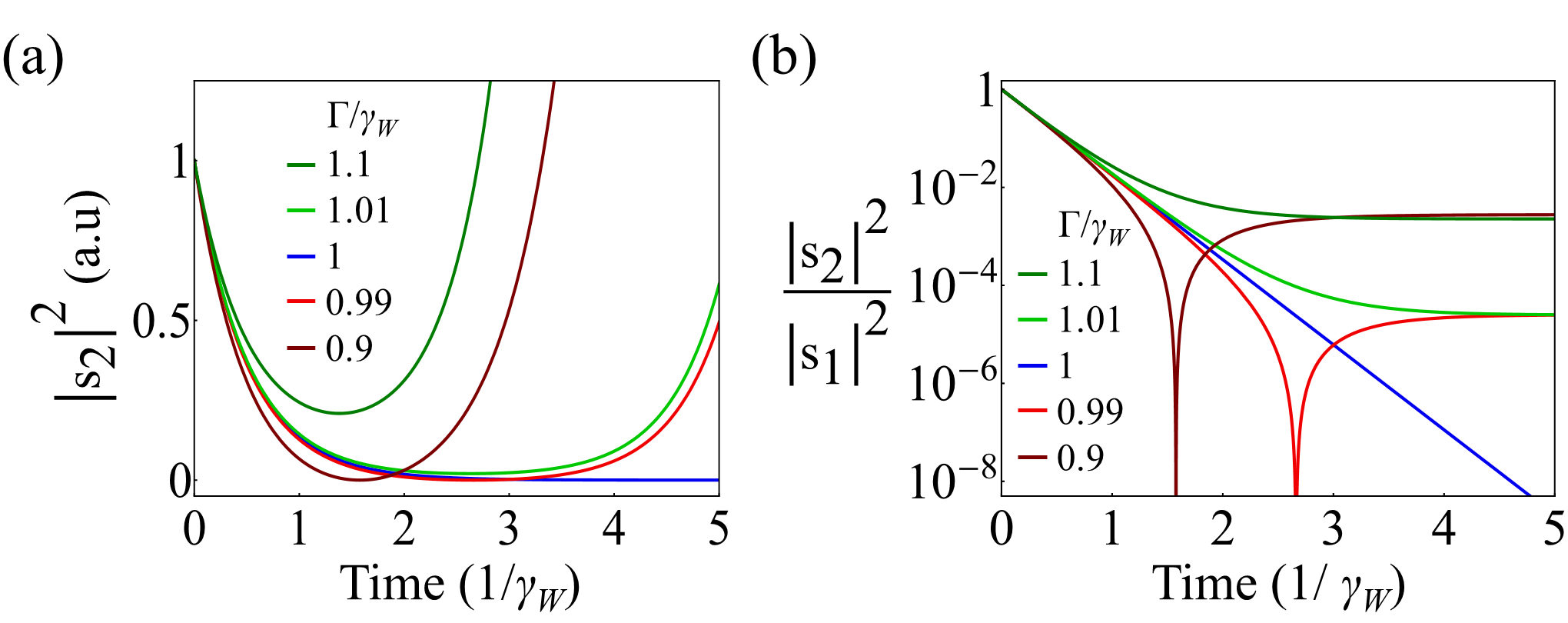}
	\caption{(a) Plots of the output optical power $|s_2|^2 $ for different values of the growth rate of the input signal $\Gamma$ deviating from its ideal value $\gamma_W$. When $\Gamma<\gamma_W$, the output power reaches zero before it starts to grow. On the other hand, for $\Gamma>\gamma_W$, the output power decreases first but never reaches zero before it starts to grow. (b) Log-scale plots of the ratio $|s_2|^2/|s_1|^2$ demonstrate that it remains small in all cases. In other words, despite the fact that the output power eventually increases with time, most of the input power is still stored in the microring.}
	\label{Fig_VaryingInput} 
\end{figure}

\textbf{Practical considerations}---We have so far studied the process of CVA under ideal conditions. Here, we consider its practical limitations focusing mainly on non-ideal temporal wave-form shaping and nonlinear effects. To investigate the former, we consider again Eq. (\ref{Eq:CMT}) in the absence of Kerr nonlinearity and gain when the growth rate of the input signal, $\Gamma$, deviates from its ideal value of $\gamma_W$. Figures \ref{Fig_VaryingInput}(a) and (b) depict the temporal behavior of $|s_2|^2$ and $|s_2|^2/|s_1|^2$ as a function of $\Gamma$. Note that in producing these results, the input power was not switched off as before. As shown in Fig. \ref{Fig_VaryingInput}, when $\Gamma$ changes from $0.9\gamma_W$ to $1.1\gamma_W$, $|s_2|^2$ itself will first decrease before it reverses course and increases at a finite time, eventually blowing up as time goes to infinity. On the other hand, the ratio $|s_2|^2/|s_1|^2$ will always remain small. These results indicate that while the output does not vanish anymore, most of the input energy is still stored inside the microring. 

Finally, we investigate the limitations imposed by Kerr nonlinearity on the CVA process by numerically integrating Eq. (\ref{Eq:CMT}) for different values of $r$ (which is equivalent to changing the input power levels). To isolate the effect of the nonlinearity, we use here the ideal growth rate $\gamma_W$ for the input signal. In general, the nonlinear coefficient is given by the expression $\alpha_0=n_2 c \omega_0/n_0^2 V_\text{eff}$ in $\text{J}^{-1} \text{s}^{-1}$. Here $n_2$ is the nonlinear Kerr coefficient in units of $\text{m}^2 \cdot  \text{W}^{-1}$, $n_0$ is the linear refractive index and $V_{\text{eff}}=2\pi R A_{\text{eff}}$ is the effective mode volume, as expressed in terms of the effective mode area $A_{\text{eff}}$. First, we investigate the importance of nonlinear Kerr effects in microresonators that exhibits relatively low quality factors. To do so, we consider a 3D version of the structure studied in Fig. \ref{Fig_SimulationResults}, with waveguides having cross sections of $0.5\   \upmu\text{m} \times 0.22\   \upmu\text{m}$ (typical values for silicon photonics platforms). By performing modal analysis using FDTD, we can obtain the propagation constant as a function of frequency and eventually calculate the optical group index, which is found to be $n_g \sim 4$. Together with $\kappa=0.3$, we obtain the values of the loss rate $\gamma_W=0.107 \ \text{ps}^{-1}$ and the quality factor $Q\equiv \omega_0/ 2\gamma_W \sim 5700$. From the modal analysis, we also find that the fundamental TE mode has an effective area of $A_{\text{eff}}\sim 0.067 \ \upmu \text{m}^2$ at $\lambda_0=1.55 \ \upmu \text{m}$ \cite{Agarwal_NLO}. For a ring of radius $5 \  \upmu\text{m}$,  $n_0 \sim 3.47$ and $n_2 \sim 5.5 \times 10^{-18} \ \text{m}^2 \text{W}^{-1}$ \cite{Hon2011JAP} (typical values for silicon and silicon nitride based material systems), we finally obtain $\alpha_0 \sim 7.92 \times 10^{22} \ \text{J}^{-1} \text{s}^{-1}$  \cite{Kippenberg2011MBO, Chembo2016KOF, Kippenberg2018DKS}.

\begin{figure} [!t]
	\includegraphics[width=3.4in]{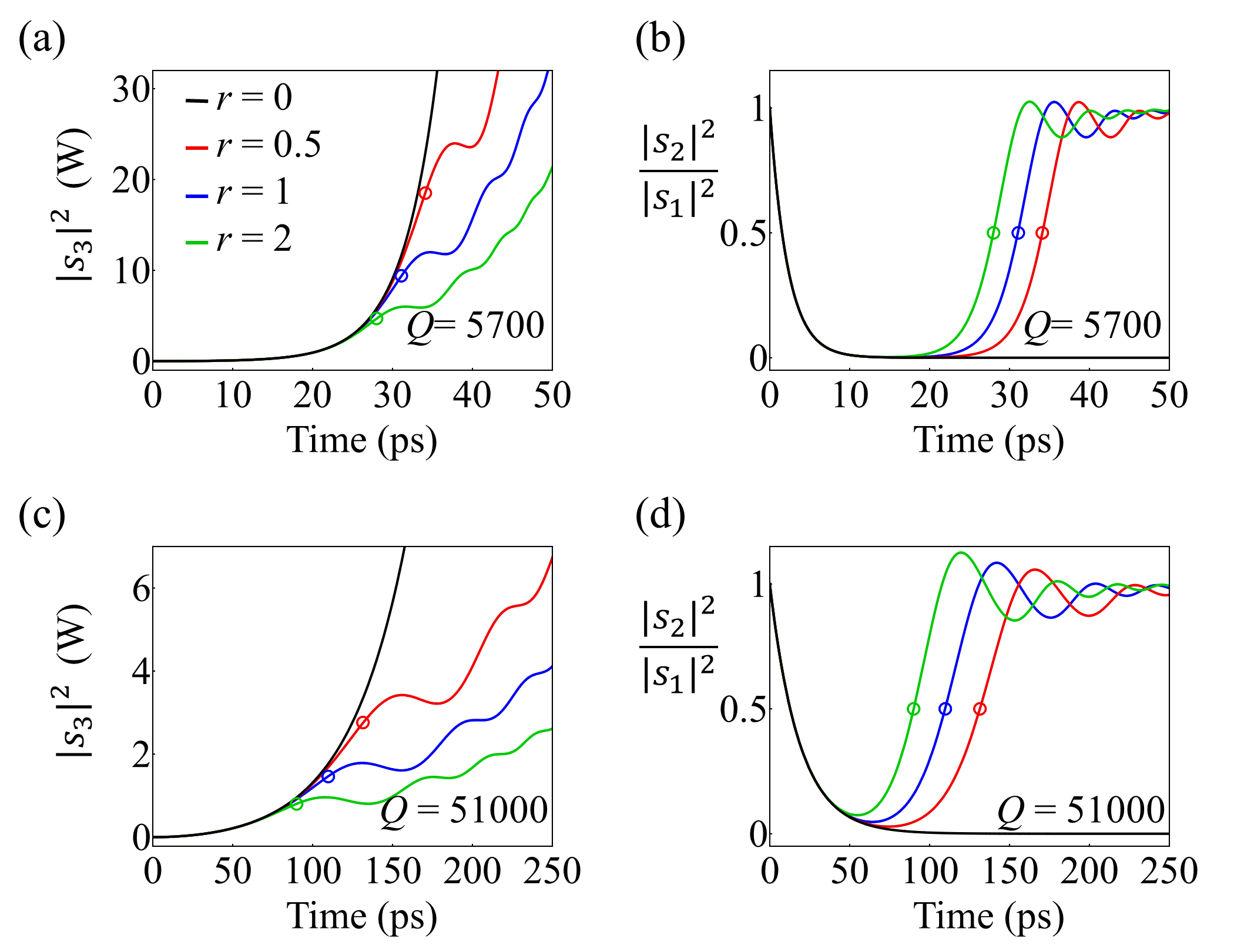}
	\caption{(a) and (b) are the interactivity power $|s_3|^2$, and the output/input power ratio $|s_2|^2/|s_1|^2$ as a function of time for different values of $r$ (equivalent to changing the input power level) for the nonlinear parameters described in the text when the resonator has a quality factor $Q \sim 5700$. (c) and (d) are similar plots when $Q \sim 51000$. In the first case, the nonlinear effects become important only at very large power values and can be thus safely neglected under typical experimental conditions. In the second scenario, however, nonlinear effects can play an important role at input power values of tens of milliwatts. Here the nonlinearity is assumed to be of the Kerr type. For other types of nonlinearities, such as thermal effects, the power levels at which the process of CVA is destroyed can be in the order of few millimeters \cite{Lipson2004OL}.}
	\label{Fig_Nonlinear} 
\end{figure} 

Figure \ref{Fig_Nonlinear} plot the optical power in the microresonators $|s_3|^2$ and the power transfer function between the output/input ports $|s_2|^2/|s_1|^2$ for different value of $r$. In these simulations, the initial input power at $t=0$ was taken to be $|s_1(t=0)|^2=1 \  \text{mW}$. From the plots, we find that the nonlinear effects eventually kick in after a transient time and `terminate' the process of CVA, as can be seen from Fig. \ref{Fig_Nonlinear}(b). This can be intuitively understood by noting that the Kerr nonlinearity results in a shift in the resonant frequency of the microring resonator which in turn destroys the delicate destructive interference at the output port. However, for the particular chosen parameters, this occurs for power values around $10 \ \text{W}$ inside the ring, corresponding to an input power of $\sim 1.1 \ \text{W}$. These are very large numbers that exceed typical experimental values. Thus, from a practical perspective, nonlinear effects will not play an important role in this specific system. If we chose different parameters, such as reducing the coupling coefficient between the waveguide and the ring to $\kappa=0.1$, which results in $Q \sim 51000$, we find that the nonlinear effects become important at power levels of $ \sim 1 \ \text{W}$ in the ring (or $\sim \ 20 \text{mW}$ in the input waveguide) as can be seen in Figs. \ref{Fig_Nonlinear}(c) and \ref{Fig_Nonlinear}(d). Here the initial input power in the simulations was taken to be $1 \ \text{mW}$. These are rather realistic values, although they remain large \cite{Lipson2004OL}. If one considers other nonlinear effects such as for example thermal nonlinearities, these power levels can be considerably lower at the level of a few milliwatts \cite{Lipson2004OL}. Our analysis thus indicates that nonlinear effects can pose a serious challenge for implementing CVA for resonators having high-quality factors and large nonlinear effects.    

\section{Conclusion}
In summary, we have investigated CVA in the context of microring resonators. We have presented an analytical expression describing CVA in microring resonators and used it to make a connection with the process of CPA. We have also presented an intuitive description of CVA using cascaded beam splitters, highlighting the interferometric nature of the effect. Importantly, we have utilized 2D full-wave simulations using FDTD to demonstrate the feasibility of CVA in microring resonators based on silicon platforms. Finally, we have investigated some of the practical limitations arising from non-ideal wavefronts and nonlinear effects. In this context it will be of particular interest to investigate similar schemes in connection with frequency comb devices \cite{Ferdous2011NP,Brasch2015S} and soliton crystals \cite{Cole2017NP} for nonlinear optical applications. We plan to do this in future works. Our results, although presented in the framework of photonics, are also relevant to other wave-physics contexts like matter waves, RF and acoustics.

%\appendix
%\section{}
%\appendix
%\section{Appendix 1}

\begin{acknowledgments}
R.E. acknowledges support from ARO (Grant No. W911NF-17-1-0481), NSF (Grant No. ECCS 1807552), and the Max Planck Institute for the Physics of Complex Systems. T.K. acknowledges partial support from ONR (Grant No. N00014-19-1-2480) and NSF (Grant No.  EFMA1641109).  
\end{acknowledgments}

Q.Z. and L.S. contributed equally to this work.


\bibliography{Reference}

\begin{thebibliography}{60}
\bibitem [{1()}]{1}%
  \BibitemOpen
  \href@noop {} {}\bibinfo {note} {The terminology 'virtual' should not be
  confused with the notion of virtual photons.}\BibitemShut {Stop}%
\end{thebibliography}

\end{document}